# Comment on - Synthesis, growth and characterization of a new metal–organic NLO material: Dibromo bis (*L*-proline) Cd(II) [J. Mol. Struct. 1080 (2015) 37–43]

Bikshandarkoil R. Srinivasan, Madhavi Z. Naik, Kedar U. Narvekar
Department of Chemistry, Goa University, Goa 403206, India
Email: srini@unigoa.ac.in Telephone: 0091-(0)832-6519316; Fax: 0091-(0)832-2451184

**Abstract**

The title paper reports a study on the crystal growth and characterization of dibromobis(*L*-proline)cadmium(II) **1**. Many points of criticism, concerning the crystal structure of **1** and the magnetic properties of **1**, dibromobis(*L*-proline)zinc(II) **2** (J. Mol. Struc. 1033 (2013) 121–126) and diiodobis(2-aminopyridine)cadmium(II) **3** (J. Mol. Struc. 1042 (2013) 25–31) are reported to show that compounds **1** to **3** are not soft magnets but instead diamagnetic solids.

**Comment**

The authors of the title paper [1] report on the crystal growth and characterization of dibromobis(*L*-proline)cadmium(II) **1** which according to them is supposed to possess soft magnetic properties at room temperature. The same group has earlier reported that dibromobis(*L*-proline)zinc(II) **2** and diiodobis(2-aminopyridine)cadmium(II) **3** are soft magnets [2, 3]. As it is well known that compounds of Zn and Cd are diamagnetic, these reports attracted our attention and were taken up for scrutiny.

The authors report to have grown crystals of **1** from an aqueous solution containing *L*-proline and CdBr2 in 2:1 mole ratio by the slow evaporation method. The crystal structure of **1** reveals tetra coordination around Cd(II). The central metal is bound to two crystallographically independent bromide ligands and to two unique monodentate *L*-proline ligands via the carboxylate oxygen resulting in a discrete tetrahedral {$CdBr_2O_2$} complex isostructural with dichlorobis(*L*-proline)zinc(II) reported by Yukawa et al in 1985 [4]. Disregarding this, the authors of [1] reported "*The complex has a very similar structure to that of CdCl$_2$ (Hpro); it consists of a one dimensional polymer bridged by bromine atoms and carboxyl oxygen atoms.*"

In [CdCl$_2$(Hhpro)] (Hhpro = 4-hydroxy-*L*-proline) reported by Yukawa et al [5] the central Cd(II) is hexacoordinated with both the chloride ligands as well as the *L*-proline functioning as μ$_2$-bridging bidentate ligands. The Yukawa group has also reported another hexacoordinated Cd(II) compound, namely dichloro(*L*-proline)cadmium(II) hydrate [6] which is a one dimensional polymer. Without taking into account that the hexacoordinated Cd(II) compounds exhibit a Cd:*L*-proline ratio of 1:1 unlike the bis(*L*-proline) compound **1** and the structural features of the hexacoordinated Cd(II) compounds and the tetracoordinated **1** are quite different [7] the authors reported, '*The metal ligand coordination and molecular geometry are similar to that observed in related amino acid containing compounds [11], [12] and [13]*' where citations 11 and 12 are the papers of Yukawa on the hexacoordinated Cd(II) compounds [5, 6].

It is well known that the salt CdBr$_2$ (or ZnBr$_2$) and the amino acid *L*-proline are diamagnetic in nature and one does not normally expect any new magnetic properties for a compound which is a chemical combination of CdBr$_2$ (or ZnBr$_2$) and *L*-proline. There are also no reports in the literature on Cd(II)-amino acid compounds showing they are not diamagnetic. However, in the discussion of the magnetic study of **1** authors reported '*The Zn (II), Cd (II) metal complexes possess diamagnetic properties but some metal coordinated complexes may possess different properties*'. Based on such a reasoning and the measurements performed using a vibrating sample magnetometer (VSM) on a crystalline sample of **1** at room temperature (RT), authors declared it as a soft magnet. Although the magnetization curve reported for **1** is a straight line passing through the origin typical of diamagnetic behaviour, the authors claim that they have obtained a hysteresis loop for soft materials, due to a lack of expertise to interpret magnetic data. It is not clear how the authors calculated the saturation magnetization (Ms) as $558.12 \times 10^{-6}$ emu/g from this graph, and determined a remarkable value of $368.58 \times 10^{-9}$ emu/g for the remnant magnetization (Mr). We opine that this is an example of making an incorrect claim of RT soft magnetic behaviour, due to being unaware of the errors associated with measurement of magnetic signals, a point well described recently by Garcia et al [7].

For the soft magnetic behaviour of **1** as well as **2** the following identical reasoning "*The coordination environments are changed from tetrahedral to distorted tetrahedral structure (diamagnetic to soft magnetic nature). The weak interactions of N-H···O, N-H···Br hydrogen bondings are also responsible for*

*such kind of magnetic properties"* is given in both the papers. We do not agree with this because the distorted structure is a consequence of the dissimilar Cd-Br (or Zn-Br) and Cd-O (or Zn-O) bond distances in **1** (or **2**). The authors are unaware that the presence of H-donors and H-acceptors in the crystal structure are responsible for the weak interactions and several metal free amino acid compounds which exhibit N-H⋯O, N-H⋯Br interactions are well documented in the literature [8]. The above arguments are equally valid for diiodobis(2-aminopyridine)cadmium(II) **3** [3] which is claimed as soft magnet by the authors based on a very strange magnetization curve and the reasoning that weak N-H⋯O and N-H⋯I interactions in **3** are responsible for the magnetic properties. The diamagnetic nature of all compounds can be evidenced from the reported NMR spectra in the commented papers, which exhibit sharp signals. In summary, compounds **1**-**3** are not soft magnets.